\documentclass[useAMS,usenatbib]{mn2e}
\usepackage{graphicx}
\usepackage{epsfig}
\usepackage{rotating}

\newcommand{\HII}{H\,{\sc ii}}
\newcommand{\SII}{[S\,{\sc ii}]}

\newcommand{\OIII}{[O\,{\sc iii}]}
\newcommand{\OII}{[O\,{\sc ii}]}
\newcommand{\OI}{[O\,{\sc i}]}

\newcommand{\NII}{[N\,{\sc ii}]}

\def\SNR{\mbox{{MCSNR J0512$-$6707}}}

\def\p0{\phantom{0}}

\title[Optical discovery of an LMC SNR]{Optical discovery and multiwavelength investigation of supernova remnant MCSNR\,J0512--\,6707 in the Large Magellanic Cloud}
\author[Warren Reid, et al.]{Warren A. Reid$^{1,2,4}$\thanks{E-mail:
warren.reid@outlook.com (WR); } Milorad Stupar $^{4}$\thanks{E-mail:
m.stupar@uws.edu.au (MS); } Luke M.~Bozzetto${^4}$ Q. A.
Parker$^{3,5}$\thanks{E-mail: quentin.parker@mq.edu.au (QAP)}
\newauthor M. D.~Filipovi\'c${^4}$\thanks{E-mail: m.filipovic@uws.edu.au} \\
$^{1}$Department of Physics and Astronomy, Macquarie University, Sydney, NSW 2109, Australia\\
$^{2}$Centre for Astronomy, Astrophysics and Astrophotonics, Macquarie University, Sydney, NSW 2109, Australia\\
$^{3}$Australian Astronomical Observatory, PO Box 296, Epping, NSW 1710\\
$^{4}$University of Western Sydney, Locked Bag 1797, Penrith South DC, NSW 1797, Australia\\
$^{5}$Department of Physics, University of Hong Kong, Pokfulam, Hong Kong}

\begin{document}

\date{Accepted 2014 December 20. Received 2014 December 20; in original form 2014 December 20}

\pagerange{\pageref{firstpage}--\pageref{lastpage}} \pubyear{2002}

\maketitle

\label{firstpage}

\begin{abstract}
We present optical, radio and X-ray data that confirm a new supernova remnant (SNR) in the Large Magellanic Cloud (LMC) discovered using our deep H$\alpha$ imagery. Optically, the new SNR~has a somewhat filamentary morphology and a diameter of 56~$\times$~64 arcsec (13.5~$\times$~15.5\,pc at the 49.9\,kpc distance of the LMC). Spectroscopic follow-up of multiple regions show high \SII/H$\alpha$ emission-line ratios ranging from 0.66\,$\pm$0.02 to 0.93\,$\pm$0.01, all of which are typical of an SNR. We found radio counterparts for this object using our new Australia Telescope Compact Array (ATCA) 6\,cm pointed observations as well as a number of available radio surveys at 8\,640 MHz, 4\,850 MHz, 1\,377 MHz and 843 MHz. With these combined data we provide a spectral index $\alpha$\,$\approx$\,--0.5 between 843 and 8\,640 MHz. Both spectral line analysis and the magnetic field strength, ranging from $124 - 184~\mu$G, suggest a dynamical age between $\sim$2,200 and $\sim$4,700\,yrs. The SNR has a previously catalogued X-ray counterpart listed as HP\,483 in the ROSAT Position Sensitive Proportional Counter (PSPC) catalogue.
\end{abstract}


\section{Introduction}

SNRs represent a prime source of heavy elements which are synthesised in their high mass progenitors. They therefore play a major role in the enrichment of their host galaxies. Supernova (SN) explosions produce shocks, powerful blast waves and a radiation front strong enough for the resulting SNRs to shape and heat the interstellar medium (ISM), impacting on the ionisation balance of the entire area in which they reside. SNRs usually produce radiation across the whole electromagnetic spectrum with detections in the optical, infrared (Laki\'{c}evi\'{c} et al. 2015), radio (due to non-thermal radiation e.g. Stupar et al. 2007), X-ray (detected by satellites e.g. Aschenbach 2002), the region of soft gamma-rays (due to the presence of residual young neutron stars) and even in the ultraviolet (Blair et al. 2006). Multi-wavelength studies of SNRs provide information on several crucial aspects of stellar evolution and the physical behaviour of shocked material. They have a major influence on local magnetic fields whilst the shock waves may trigger new star formation and accelerate cosmic rays.

 Within SNRs there is a correlation between H$\alpha$ and radio-continuum emission even though the detailed fine structures do not overlap very often (see Cram, Green \& Bock, 1998). If the SN blasts into a hot, low-density ISM, H$\alpha$~emission may be faint but if the ISM is cool and dense, H$\alpha$~radiation is usually detected (with an intensity dependent on the density and temperature at the interface) revealing the shape and density of the shock front in the form of narrow filaments, elongated arc structures, bubble-like features and knots. These same structures can usually be seen in the forbidden line emission from \SII\,6717/6731 and to a more varying degree in \NII\,6548/6584, \OI\,6300/6364, \OIII\,4959/5007 and the \OII~doublets close to 3727\,\AA~and 7325\,\AA. Following Fesen, Blair \& Kirshner (1985), a ratio of \SII~to H$\alpha$ $>$~0.5 implies a shock front indicative of an SNR as does strong \OI~and \OII~lines plus the supporting presence of \NII~(Sabbadin et al. 1977). In addition, the detection of H$\alpha$, rather than \OIII,~leading the radiation front would support the rejection of any Wolf-Rayet (WR) shell scenario (Gruendl et al. 2000).

At a distance of $\sim$49.9\,kpc (Reid \& Parker 2010 and references therein) the LMC is an ideal laboratory to study the different varieties of SNRs in great detail across an entire galaxy. The LMC is close enough that SNRs are resolved sufficiently for multiwavelength studies to be conducted. Since SNRs yield key information on the death rate of high mass stars and the number of white dwarfs exploding as Type Ia SN, an improved inventory of LMC SNRs, complete as possible, is required to put constraints on these values and improve our understanding of the final stages of stellar evolution. The LMC SNR sample is currently estimated at 59 remnants\footnote{estimated by the authors using 31 SNRs from Chandra catalog (hea-www.cfa.harvard.edu), 54 from Badenes et al. (2010) and recent discoveries including those from Maggi et al. (2014), Bozzetto et al. (2014a,b), Warth et al. (2014) and Kavanagh et al. (2015)}.


The deep, stacked UKST H$\alpha$ survey used to search for planetary nebulae in the LMC (Reid \& Parker,
2006a, 2006b, 2013), also revealed several candidate SNRs, the first of which we report here, thanks to the
optical depth of the H$\alpha$~survey to R$_{equiv}$~mag 22, and the method of
image analysis. Spectroscopic followup of
these objects (originally identified as SNR candidates in 2004) using 2dF on the Anglo-Australian Telescope (AAT) and the 1.9-m telescope at the South African Astronomical Observatory (SAAO) have permitted us to confirm identifications and conduct an optical spectroscopic and multiwavelength analysis.

\begin{figure}
  \includegraphics[width=0.48\textwidth,angle=0]{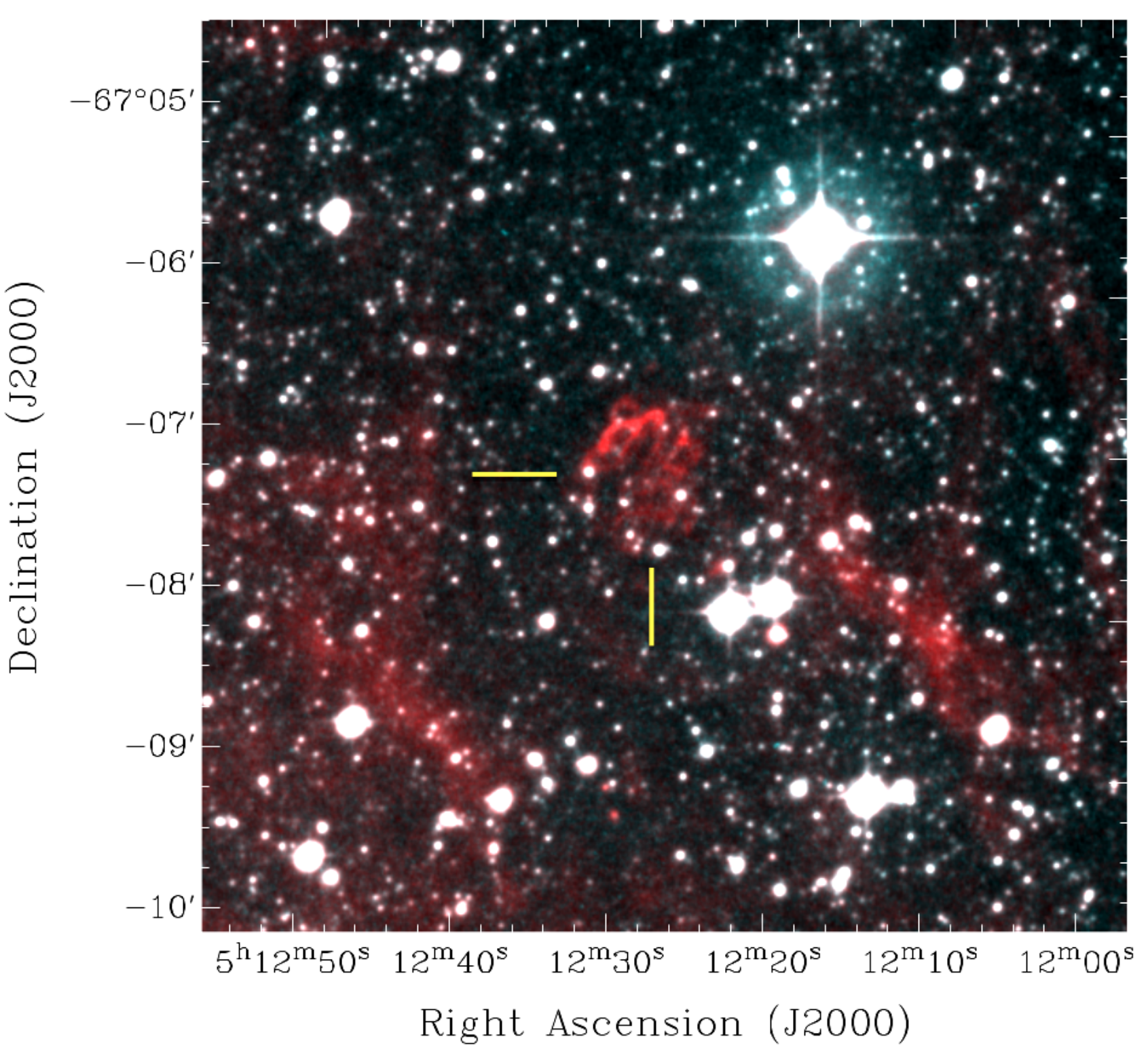}\\
  \caption{The median stacked and combined H$\alpha$ and broad band red optical image of RP1577 where H$\alpha$~is shown as red and the short red is shown as white. }\label{fig2}
\end{figure}

The rest of the paper is structured as follows. In section~\ref{section2} we present the optical observational data including images and spectra. In subsection~\ref{section2.1} we derive an estimate for the expansion velocity and age of the SNR and in~\ref{section2.2} estimate the radial velocity from key emission lines. In section~\ref{section3} we provide supporting radio data, the results of infrared imaging and previous X-ray observations. Section~\ref{section4} is our conclusion where we summarize the combined multiwavelength data used to achieve our classification.

\label{section1}

\section{Optical observations and discovery}
\label{section2}


The new SNR candidate, given the Reid \& Parker reference name RP1577 and presented here for the first time, was discovered using deep-stack H$\alpha$~and short red images of the
LMC, SMC and surrounding regions (Reid \& Parker 2006a). An H$\alpha$ interference filter of exceptional
specification and quality (Parker \&
Bland-Hawthorn 1998) was used to cover a substantial fraction of the
UKST's large field. Its central wavelength (6590\,\AA) and
bandpass (70\,\AA~FWHM) work effectively in the UKST's fast f/2.48
converging beam. With a peak transmission of $>$85 per cent, the H$\alpha$ filter is also partially sensitive to
\NII6583\,\AA. Overlapping fields with non-standard 4-degree centres enabled
full contiguous coverage in H$\alpha$ despite the circular
aperture. High resolution, panchromatic Tech-Pan film with peak sensitivity at H$\alpha$ was used as the photographic detector (see Parker et al. 2005 for more details).

\begin{table*}
\caption{Observing Logs for the follow-up of emission object and SNR candidate RP1577 in the LMC. The first column gives the telescope name, second gives the instrument name, third gives the observation date, the fourth gives the grating name, the fifth gives the dispersion in angstroms, the sixth gives the central wavelength, the seventh gives the wavelength coverage in angstroms, the eighth gives the exposure time in seconds, the ninth gives the number of exposures taken.}
\begin{tabular}{|l|c|c|c|c|c|c|c|c|c|}
  \hline \hline
   &   &    &   &    &   &    &   \\
   Telesc. & Inst. & Date & Grating  &  Dispersion & Central  & Coverage &
   T$_{exp}$ & N$_{exp}$ \\
      &  & &  Dispenser &  \AA/pixel & $\lambda$ (\AA) & $\lambda$ (\AA) & s  & \\
   \hline
    AAT &  2dF &  16 Dec-04   &   300B    &   4.3    &   5852 & 3660 - 8000 &   1200    &   3   \\
    AAT & 2dF &   18 Dec-04   &   1200R   &   1.1   &   6793 & 6220 - 7340 &   1200    &   2    \\
 1.9-m & ccd   &   7 March-14   &   300    &   5   &   5800  & 3850 - 7738  &   1800    &   2    \\
    \hline
\end{tabular}
\label{table 1}
\end{table*}

Over a period of three years, from 1997, a series of 12 repeated
narrow-band H$\alpha$ and 6 matching broad-band `SR' (Short Red)
exposures of the central LMC field were also taken (see Reid \& Parker 2006a). The purpose was that these images should be stacked to
produce a single H$\alpha$ and short red image with an overall depth
increase of 1 magnitude over a single image frame. The twelve
highest quality and well-matched UK Schmidt Telescope 2-hour
H$\alpha$ exposures and six 15-minute equivalent SR-band exposures
were selected. From these exposures, deep, homogeneous, narrow-band
H$\alpha$ and matching broad-band SR maps of the
entire central 25 deg$^{2}$ square of the LMC were constructed.


The `SuperCOSMOS' plate-measuring machine at the Royal Observatory
Edinburgh (Hambly et al. 2001) scanned, co-added and pixel
matched these exposures creating 10$\mu$m (0.67~arcsec) pixel data
which goes 1.35 and 1 magnitudes deeper than individual exposures,
achieving the full canonical Poissonian depth gain, e.g.
Bland-Hawthorn, Shopbell \& Malin (1993). This gives a depth
$\sim$21.5 for the SR images and $R_{equiv}\sim$22 for H$\alpha$
($4.5\times10^{-17}ergs~cm^{-2}~s^{-1}~$\AA$^{-1}$) which is at
least 1-magnitude deeper than the best wide-field narrow-band LMC
images previously available. An accurate world co-ordinate system
was applied to yield sub-arcsec astrometry, essential for success of
the spectroscopic follow-up observations.

The resulting H$\alpha$ and short red (SR) maps were originally overlayed with false colours in order to detect planetary nebulae (Reid \& Parker, 2006a,b,2013) but also proved an excellent
tool for uncovering a large number of hot and cool emission-line stars (Reid \& Parker, 2012). In addition to these discrete objects, the maps were able to reveal the fine morphological
characteristics of SNRs, which appear as complex interwoven streams, elongated arcs, bubbles and narrow filaments in H$\alpha$. The newly identified candidate is shown in
Fig.~\ref{fig2}~where H$\alpha$ is coloured red and the short red continuum image appears white due to both blue and green colours being assigned to the same image.
Although these structures can be closely correlated with radio emission, they usually become quite fragmented in old remnants and may extend beyond, within or along
the boundaries of the radio remnant (see Stupar \& Parker 2009). In addition to the H$\alpha$/SR combined images, we used the Magellanic Cloud Emission-Line Survey (MCELS; Smith et al. 1998) to directly compare the strength of the three narrow bands of \OIII\,5007 (blue), H$\alpha$ (red) and \SII\,6716+6731 (green). When the colour maps are evenly combined the SNR candidate becomes evident by the strong green (\SII~line) colour.

RP1577 has a central position of (J2000) RA 05$^{h}$ 12$^{m}$ 28$^{s}$ DEC -67$^{\circ}$ 07$^{\prime}$ 20$^{\prime}$$^{\prime}$ determined using SuperCOSMOS astrometry (Hambly et al 2001) and a measured angular extent from the H$\alpha$ imagery of 56 $\times$ 64 arcsec. This equates to a physical extent of 13.5 $\times$ 15.5\,pc at the LMC distance. With physical diameters ranging from 0 to 120\,pc (Badenes et al. 2010), the cumulative distribution of SNR diameters in the LMC have been found to be roughly linear (eg. Chu \& Kennicutt 1988; Bandiera \& Petruk 2010) up to a cutoff at a physical radius of $\textit{r}_{\textrm{cut}}$ $\sim$30\,pc (Badenes et al. 2010). This means that the 15.5\,pc size of RP1577 places it among the smaller 15 per cent of LMC SNRs, joining six other SNRs in the 10-20\,pc range.

\begin{table*}
\caption{Observed line intensities and flux ratios for four different spectroscopically observed positions across three areas of RP1577. Line measurements were made
using the {\tiny SPLOT} script in the {\tiny IRAF} package. Errors for individual flux line measurements are a constant $\pm$1.9$\times$10$^{-17}$ for 2dF 300B, $\pm$1.2$\times$10$^{-17}$ for 2dF 1200R,
$\pm$1.5$\times$10$^{-17}$ for AAOmega, $\pm$2$\times$10$^{-17}$ for 1.9-m measurements. Column details are provided in the text. }
 \centering
 \tiny{
  \begin{tabular}{lccccccccccc}
  \hline\hline
   Obs.  & RA & DEC &    \multicolumn{4}{c}{Flux (erg cm
   s$^{-1}$)}  &\NII/ &\SII/ & \SII\,ratio  & $\textit{n$_{e}$}$ & $\textit{T}$$_{e}$ \\
       &  (J2000) & (J2000) & H$\beta$ & \OIII$_{5007}$ & H$\alpha$ & \NII$_{6583}$ &
   H$\alpha$   &  H$\alpha$ &  6716/6731  & (cm$^{3}$) & (K) \\
    (1) & (2) & (3) & (4) & (5) & (6) & (7) & (8) & (9) & (10) &
    (11) & (12) \\
\hline
  2dF-300B & 05 12 28.4 & -67 07 29  & 2.22E-15 & 3.84E-15   & 6.95E-15 &  1.49E-15   &   0.32$\pm$0.02 & 0.46$\pm$0.02     &  1.35$\pm$0.03  &   70$\pm$9 & 35,000$\pm$10,000 \\
  2dF-1200R & 05 12 28.4 & -67 07 29  & - & -   & - &  -                              &   0.25$\pm$0.02 & 0.66$\pm$0.02     &  1.44$\pm$0.03    &   $<$26 & -    \\
  1.9-m\,1 & 05 12 31.0 & -67 06 59  & 1.81E-14 & 2.20E-14       & 6.40E-14 &  1.16E-14      & 0.24$\pm$0.01   & 0.78$\pm$0.01      & 1.40$\pm$0.02   &   26.5$\pm$8 & -\\
  1.9-m\,2 & 05 12 26.0 & -67 07 05  & 3.2E-15 & 7.17E-15        & 1.17E-14 &  2.32E-15      & 0.30$\pm$0.01   & 0.93$\pm$0.01     & 1.44$\pm$0.02 &   $<$26  & - \\
 \hline
\end{tabular}}\label{table 2}
\end{table*}

\begin{figure}
  \includegraphics[width=0.48\textwidth]{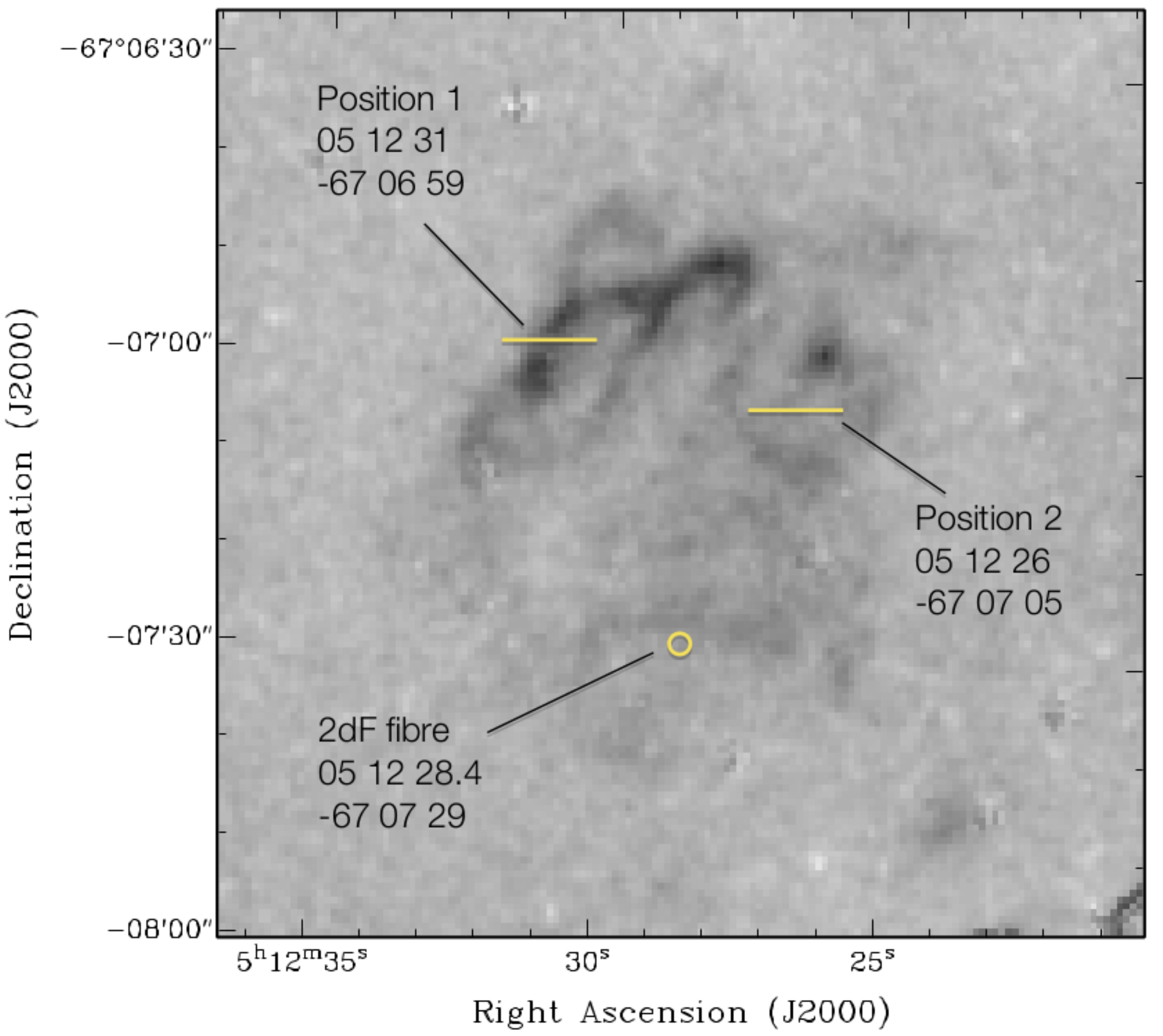}\\
  \caption{The (H$\alpha$/SR) quotient image of RP1577 showing circular loops to the north of parallel, stratified
  filaments which outline the exterior of the expanding shell. The position of the 2dF spectroscopic observation is
  shown by the yellow circle close to the center of the object. Horizontal lines mark the positions of long-slit observations.}\label{fig9}
  \includegraphics[width=0.46\textwidth]{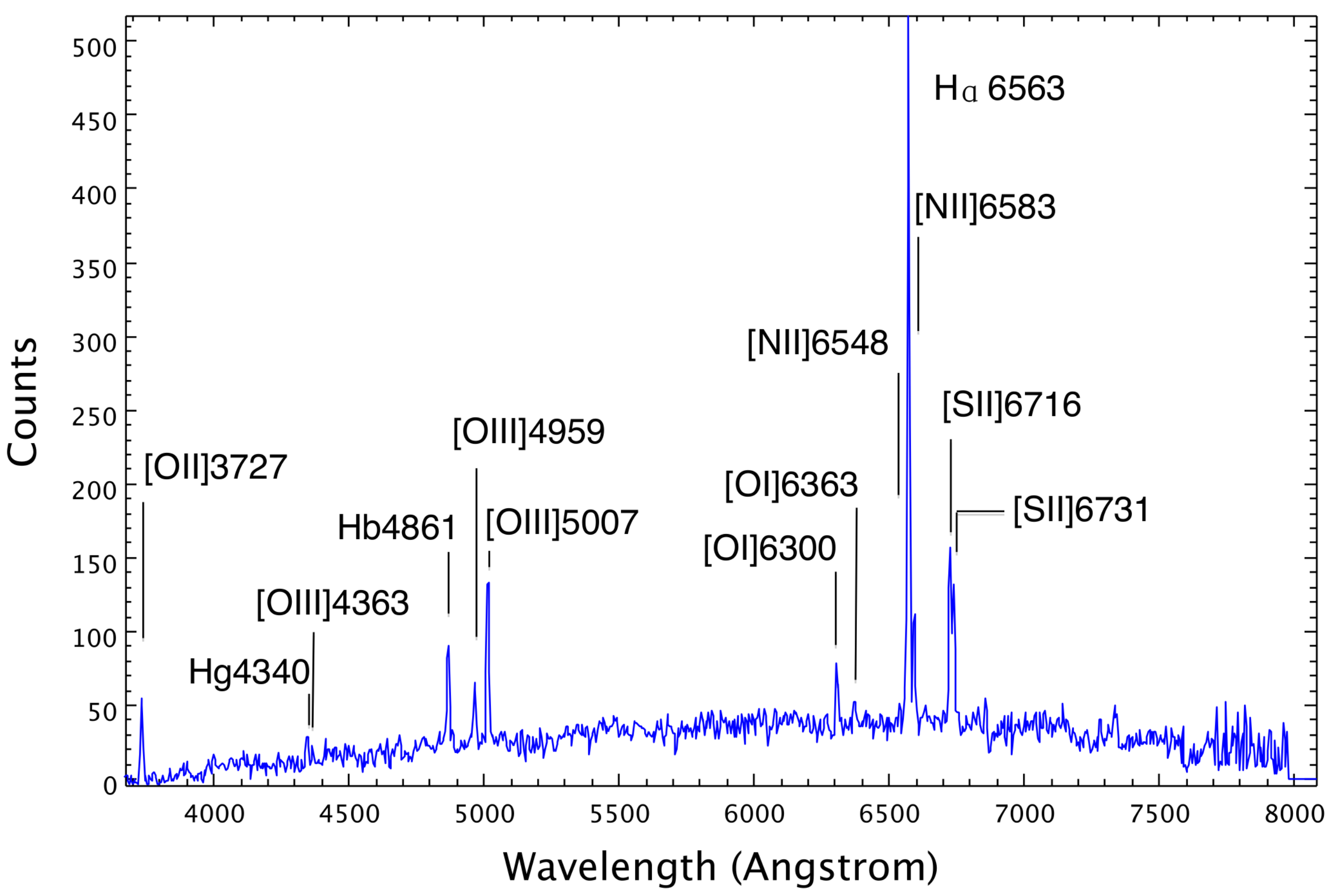}\\
  \caption{The 2dF spectroscopic observation of RP1577 indicated by the yellow circle in Figure~\ref{fig9}. The H$\alpha$/\SII~ratio is 0.46$\pm$0.02. The extinction of $\textit{c}$(H$\beta$) = 0.45 is based on the Balmer decrement.}\label{fig10}
  \end{figure}
  \begin{figure}
  \includegraphics[width=0.46\textwidth]{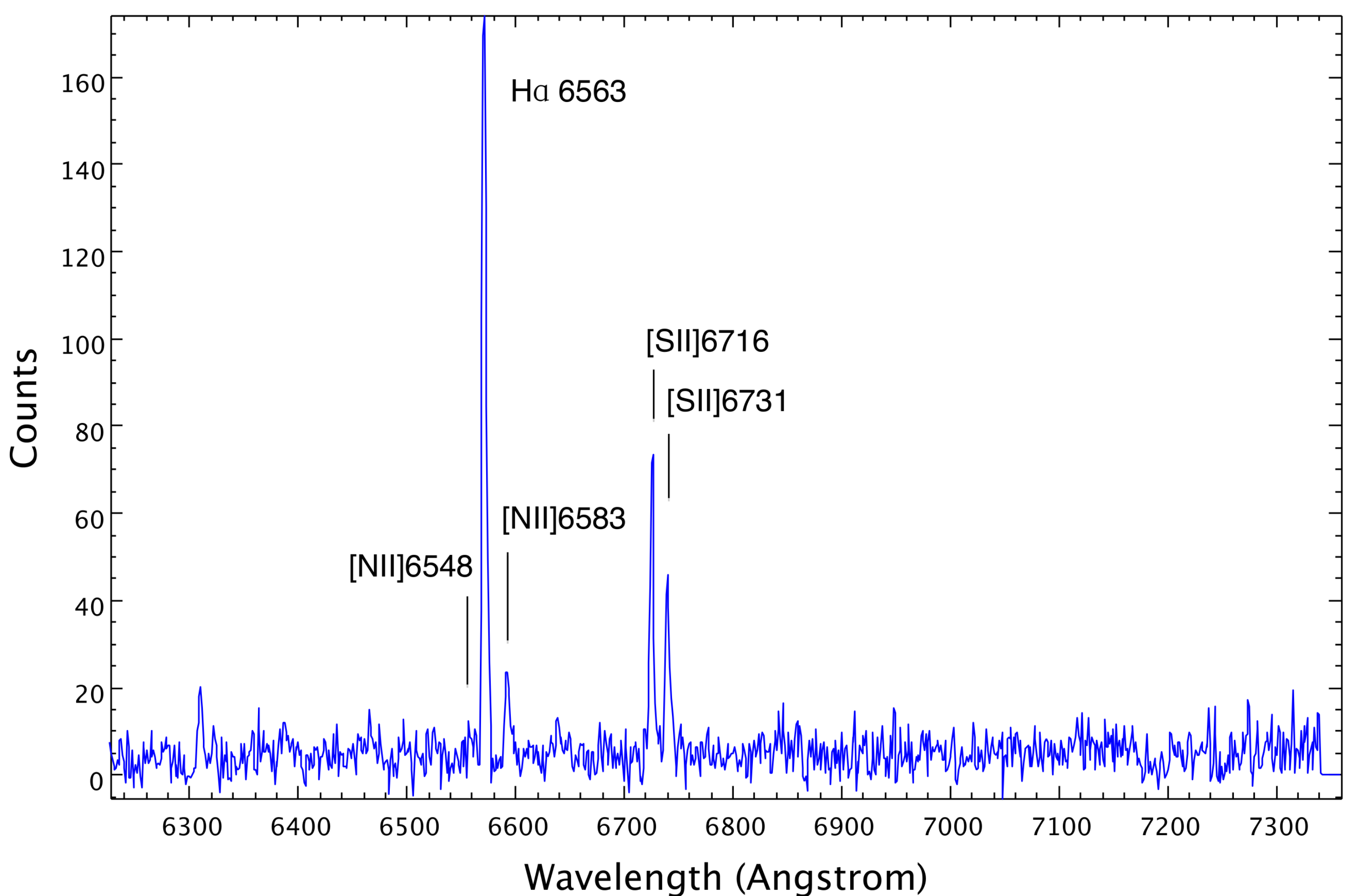}\\
  \caption{The 2dF 1200R medium resolution spectroscopic observation of RP1577 indicated by the yellow circle in Figure~\ref{fig9}. The higher resolution of this spectrum shows the weak \OI6300 \& 6363 lines. The \SII/H$\alpha$ ratio is 0.66$\pm$0.02. }\label{fig11}
\end{figure}

\begin{figure}
 \includegraphics[width=0.49\textwidth]{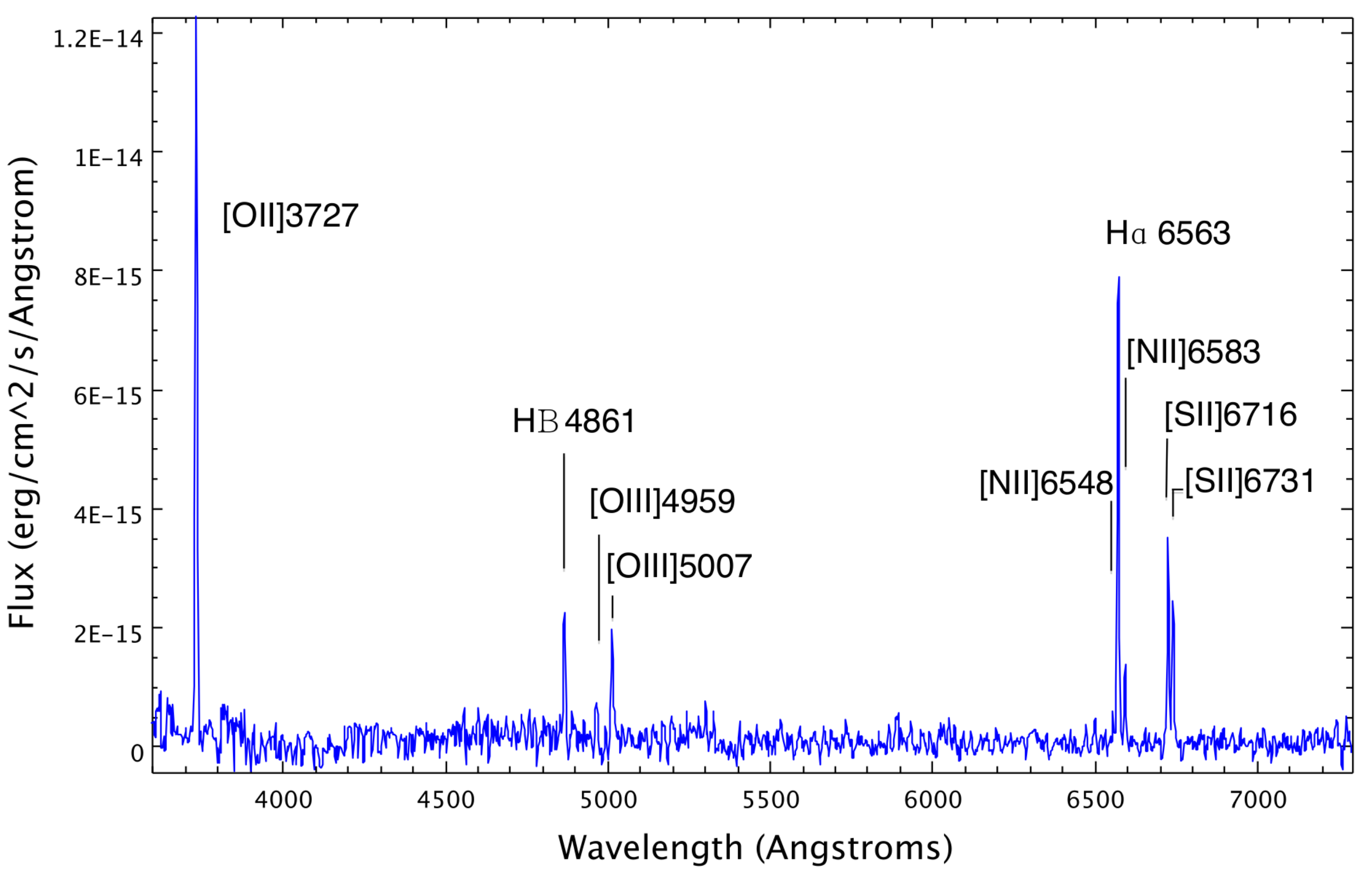}\\
  \caption{The long-slit, spectroscopic observation of RP1577 at position (J2000) RA 05$^{h}$ 12$^{m}$ 31$^{s}$ and DEC --67$^{\circ}$ 06$^{\prime}$ 59$^{\prime\prime}$. The \SII/H$\alpha$~observed ratio at this position is 0.78$\pm$0.01. The extinction of $\textit{c}$(H$\beta$) = 0.28 is based on the Balmer decrement.}\label{fig12}
 \includegraphics[width=0.51\textwidth]{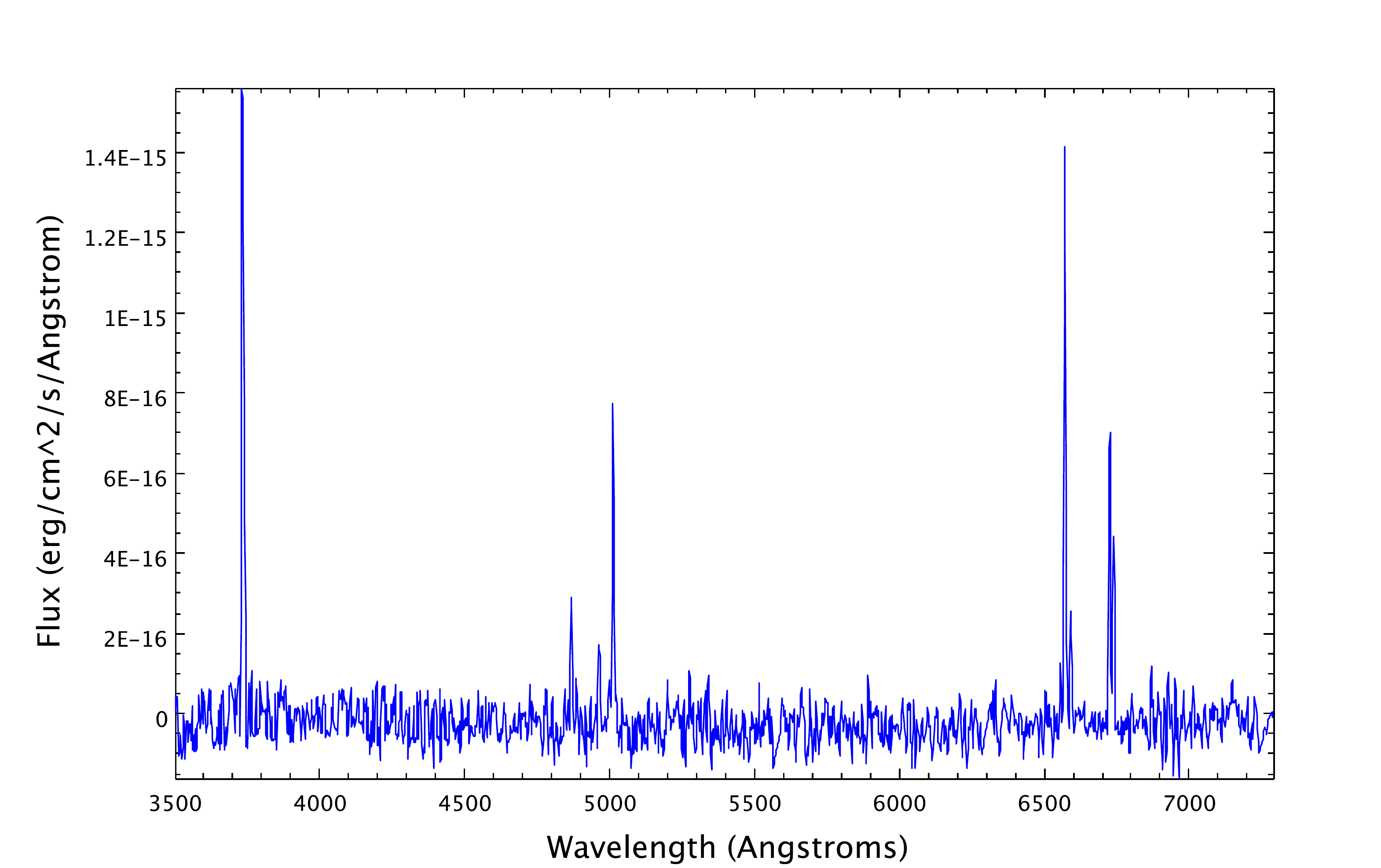}\\
  \caption{The long-slit, spectroscopic observation of RP1577 at position (J2000) RA 05$^{h}$ 12$^{m}$ 26$^{s}$ and DEC --67$^{\circ}$ 07$^{\prime}$ 05$^{\prime\prime}$. The \SII/H$\alpha$~observed ratio of 0.93$\pm$0.01 at this position is the highest ratio of all our spectroscopic observations of RP1577. The extinction of $\textit{c}$(H$\beta$) = 0.33 at this position is based on the Balmer decrement.}\label{fig13}
\end{figure}

To eliminate the possibility that RP1577 was the product of a WR shell we checked the available databases for the presence of any WR star or candidate inside (or in the vicinity of) the shell but none were found within a three arcmin radius. We then compared the expanded radius of H$\alpha$ and \OIII~emission since, according to theoretical models, the outermost \OIII~usually leads the H$\alpha$ emission in WR shells (Gruendl et al. 2000). This occurs due to a decrease in the sensitivity of H$\alpha$ at temperatures greater than 10$^{4}$\,K while \OIII~emission is unaffected by the higher temperatures. As a result, the morphologies of SNRs and WR shells should be somewhat different (see for example Cox 1972, Gruendl et al. 2000). As these models depend on the interaction of the shells with the interstellar medium, they remain as a useful diagnostic whether or not they are used to identify WR shells. Using the MCELS survey we overlayed the \OIII~and H$\alpha$ maps in order to check whether \OIII~was leading at any point around RP1577. Despite the low 2.5 arcsec/pixel resolution of the MCELS survey, a pixel by pixel comparison along the shock front showed that H$\alpha$ was in fact leading the \OIII~emission, clearly supporting the rejection of any WR shell scenario.

  In Table~\ref{table 1} we summarise details regarding the
spectroscopic follow-up observations. A five night observing run on the AAT using 2dF (Lewis et al.
2002) was undertaken in December 2004 to spectroscopically confirm LMC emission candidates. The large corrector lens incorporates an atmospheric dispersion compensator, which is essential for wide wavelength coverage using small diameter fibres. Specifically targeted long-slit spectra were obtained using the 1.9-m telescope at the SAAO in March, 2014.

Optical properties for the new SNR in
the LMC are given in Table~\ref{table 2}. Column (1) indicates the telescope and observing method used at each position. The AAT observations were conducted using the 2dF multi-fibre spectroscopy system where a each fibre covered 2 arcsecs diameter on the sky. Column (2) gives the (J2000) position of
the optical center of the observed position in right ascension. Column (3)
likewise, gives the (J2000) declination of the same position. Each observed
position was determined using the H$\alpha$ map and checked against SuperCOSMOS astrometry. A comparison of the resulting spectral signature was intended to provide a good overall analysis of the ratio of \SII~to H$\alpha$ and \NII~across the nebula, all measured with gaussian line fitting from within the {\tiny IRAF} {\tiny SPLOT} script. Column (4) gives the measured H$\beta$ flux at the position of each slit or fibre. Column (5) gives the measured flux of the \OIII\,5007 emission line at the observed position. Flux calibration of the 2dF line intensity was conducted using the method outlined in Reid \& Parker (2010). For the purpose of diagnostics the H$\alpha$ flux is provided in column (6), followed by the \NII~flux in column (7). Column (8) then gives the \NII\,6583+6548/H$\alpha$ ratio including the error based on line measurements. It should be noted that the \NII\,6548 line is quantum mechanically fixed at a 1-3 intensity ratio with the \NII\,6583 line. Column (9) provides the \SII/H$\alpha$ ratio and line measurement errors for estimating the excitation of the shock-front. Column (10) shows the ratio of \SII\,6716/6731 used for estimating the electron density.  The electron density shown in column (11) is derived from the \SII\,6716/6731 ratio derived using the {\scriptsize TEMDEN} script found in the {\scriptsize STSDAS} package from {\scriptsize IRAF}. Once the \SII~ratio reaches 1.4, representing a density of 26\,cm$^{3}$ the low density limit using this method has been reached. Column (13) gives the electron temperature, derived from the \OIII\,5007+4959/\OII\,4363 emission lines. The faint \OII\,4363 line was only detectable for the purpose of measurement at one position.

The optical H$\alpha$/H$\beta$ ratio at each observed position was used to determine the extinction
constant $\textit{c}($H$\beta$) (i.e., the logarithmic extinction at
H$\beta$) at that position. These hydrogen transitions are the strongest and easiest to accurately measure in the nebula spectrum and are fortunately located very close to the other main emission lines used for diagnostics.
The observed H$\alpha$/H$\beta$ ratio, when compared to the
recombination value of 2.86 (Aller 1984), gives a logarithmic
extinction at H$\beta$ of:

\begin{equation} \textit{c}(\textrm{H}\beta) = (log(\textrm{H}\alpha /
\textrm{H}\beta) - log(2.86)) / 0.34
\end{equation}

This estimation is based on the relationship between observed and
intrinsic intensities:

\begin{equation}
\frac{I_{obs}(\textrm{H}\alpha)}{I_{obs}(\textrm{H}\beta)} =
\frac{I_{int}(\textrm{H}\alpha)}{I_{int}(\textrm{H}\beta)}
10^{-c(\textrm{H}\beta)[f(\textrm{H}\alpha)-f(\textrm{H}\beta)]},
\end{equation}
where~[$\textit{f}$~(H$\alpha$)-$\textit{f}$~(H$\beta$)] = --0.34
from the standard interstellar extinction curve. The resulting extinction values $\textit{c}$(H$\beta$) are given in the caption below each spectrum in Figs.~\ref{fig10},~\ref{fig12}~and~\ref{fig13}.

The \SII\,6717/6731 ratio of 1.35\,$\pm$0.07 from the 2dF 300B exposures results in an electron density value near 70\,$\pm$10 cm$^{3}$. In Fig.~\ref{fig9} we show the (H$\alpha$/SR) quotient image which effectively removes stars from the area. The yellow circle in the image marks the position observed with 2dF. This is approximately 10 arcsec south of the optical centre as indicated by the elliptical shape of the diffuse H$\alpha$ emission. Spectroscopic results from this position are shown in Fig.~\ref{fig10} for the 300B exposure and Fig.~\ref{fig11}~for the 1200R exposure.

Both 2dF spectra show characteristics and ratios that are quite typical of many other SNRs we have observed within the LMC. The \NII/H$\alpha$ and \SII6716/6731 line ratios from table~\ref{table 2} are in full agreement with those of other LMC SNRs (Payne et al. 2008). At the position observed by 2dF the \SII/H$\alpha$ ratio of 0.46\,$\pm$0.02 from the 300B grating is commonly expected for an SNR (Fesen et al. 1985). Higher \SII/H$\alpha$ ratios of 0.78 and 0.93 were obtained from the 1.9-m spectra in the northern regions of the object. In these positions our image data show dense filaments of emission which can be created in an SNR by post-shock plasma. Just below the centre of the SNR, the shocks have a different effect, increasing the electron temperature and ionisation structure of the recombination region, as also seen in the more prominent \OI\,6300 lines in Figs.~\ref{fig10} and~\ref{fig11}. At the adiabatic stage in the evolution of the SNR we would expect the photoionisation fields to be rather weak close to the centre where photons are re-ionising gas which is seeping back into the shock. This probably explains why the \OI\,6300 line is higher in the central region but lower in the outer shells where the \SII/H$\alpha$ ratios appear to be higher. The \OI\,6300 lines in the raw 1.9-m long-slit spectra are wide near the continuum therefore their low level in the reduced spectra (Figs.~\ref{fig12} and~\ref{fig13}) may also in part be due to sky subtraction where the combined nebulae and telluric \OI\,6300 lines span the width of the slit. Morphologically, although the overall shape is somewhat elliptical, the interior detail of the SNR shows a series of elongated filaments, which become more dense and finally curved in the form of loops to the north. With such strong spectroscopic confirmation, we now assign RP1577 the name \SNR~which includes the basic coordinates as recommended by the IAU.

\subsection{Expansion velocity and age}
\label{section2.1}

To estimate the expansion velocity and the dynamical age, we first assume that the SNR is currently in the Sedov-Taylor phase (Sedov, 1959) where the swept-up mass is large compared to the ejected mass, the shock velocity is decelerating and the evolution is approximately adiabatic. The dynamical age is then t = $\textit{C}$ $\cdot$ R\,/\,$\textit{V}_{s}$, with $\textit{C}$ as a constant which depends on the expanding velocity of shock wave after the SN explosion. This method, often used for shell-type remnants in the adiabatic stage would require $\textit{C}$ to be a value of 2/5. In this scenario~\SNR~is still in the Sedov-Taylor phase, which is typical for SNRs up to $\sim$30,000\,yrs of age and diameters up to $\sim$30\,pc (Badenes et al. 2010). As such the radii should expand as

\begin{equation}
r \sim E_{0}^{1/5} \rho^{-1/5}t^{2/5}
\end{equation}
where E$_{0}$ is the kinetic energy of the original SN, $\rho$ is the ambient gas density and $\textit{t}$ is time. The shock velocity decelerates according to

\begin{equation}
{V}_{s} = \frac{\textrm{d}r}{\textrm{d}t} \sim E_{0}^{1/5} \rho^{-1/5}t^{-3/5}
\end{equation}

Our adopted ambient initial gas density is very difficult to estimate given the location of the SNR very close to a dense \HII~region and surrounded by dense dust emission (see section 3.3). Estimates may be complicated by inhomogeneous regions of nebula with varying optical thickness or optically thick cloudlets of dust.  For a close approximation to the average initial H density, we adopt the 1 cm$^{-3}$ ambient gas density found for use with LMC SNRs by Berhuijsen (1987). This value may be compared to the initial H density value of 10\,cm$^{-3}$ for \HII~regions in the LMC as found by Pellegrini et al (2011) which takes into account gas-phase abundances and the gas-to dust ratio averaged across the densest regions. To estimate an error in the ambient gas density we allow for a rise of up to 1.2\,cm$^{-3}$ due to dense dust in the immediate area. This was estimated by measuring diffuse emission in the general 30 arcmin radius surrounding \SNR~using our 2dF fibres. The \SII\,6716,6731 lines provide electron densities (n$_{e}$) which are proportional to the square root of the emission-line volume emissivity. These measurements provide a clue as to whether the ambient environment of \SNR~is proportionally more dense than average. The electron density n$_{e}$ of ambient \HII~is between 55\,cm$^{-3}$ and the low density limit of = 26.0\,cm$^{-3}$ which is slightly higher than the standard LMC measurement of faint, diffuse emission ($<$\,26.0\,cm$^{-3}$). Since we find that diffuse \HII~regions have a low density n$_{e}$ = $\sim$600\,cm$^{-3}$, and a H density of 7 cm$^{-3}$ (see above), an n$_{e}$\,$\sim$\,38\,cm$^{-3}$ on the same scale would be equivalent to a density of $\sim$1.2\,cm$^{-3}$. For the lower error estimate we adopt a value of 0.2\,cm$^{-3}$ since a value of 0.3\,cm$^{-3}$ was found by plotting the SNR on the correlation provided by Berezhko \& V\"olk (2004).

The upper error for our adopted energy is estimated from the radio observations for this remnant (see section~\ref{section3}) and based on a surface brightness of 1.57 $\times$ 10$^{-20}$ W m$^{2}$ Hz$^{-1}$ sr$^{-1}$ at a frequency of 1 GHz, placing the object at a position on the $\Sigma$ - $\emph{D}$ diagram by Berezhko \& V\"olk (2004) equivalent to an energy ($\textsl{E}$$_{0}$) of $\sim0.25\times10^{51}$\,erg. We adopt this estimated value as it represents an upper estimate for such an SNR in the LMC. For our lower error we adopt a value of 9.5$\times10$$^{51}$\,erg, which is just below the canonical 10$^{51}$ erg value for SNRs. The radius of 7.75\,pc is found from the maximum diameter given above. This method provides an estimated dynamical age of the remnant between $\sim$2,000\,yrs and $\sim$5,000\,yrs. The method assumes that the swept-up mass will become proportionally larger than the ejected mass and the velocity will decelerate accordingly. Using the same equations, the expansion velocity may lie between $\sim$\,600 and 1400\,km s$^{-1}$.


\subsection{Radial velocity}
\label{section2.2}

The heliocentric radial velocities at each observed position of the SNR are shown in Table~\ref{table3a}. Emission lines were measured using the IRAF splot and EMSAO tasks against standard rest velocities. Column 1 gives the name of the observation, which corresponds to the positions marked in Fig.~\ref{fig9}. Columns 2 to 5 then present the heliocentric radial velocities in km s$^{-1}$ for the \OIII\,5007, H$\alpha$, \NII\,6583 and \SII\,6716 emission lines. The estimated velocity error, given in column 6 is a standard systematic error for all velocities measured on a particular observation. It was determined according to the size of the pixels, related to the resolution of the individual spectrum. The rest velocities for each element are provided in the column headers.

As measured through optical imagery in section 2, above, the \OIII~emission is less affected by temperatures greater than 10$^{4}$K and so, can be seen to move at a
slower velocity than H$\alpha$ which leads the shock front. This effect can be measured, even though expansion at positions 1 \& 2 are directionally near to the plane of the sky. Our velocity measurements show that this is true of the \OIII~line at both positions observed by the SAAO 1.9-m telescope. Velocities on the forbidden lines of \NII~and \SII~are in every case higher than those
found for H$\alpha$. This is to be expected for \SII~since the bulk abundance of this element is produced in the SN explosion.



\begin{table}
\caption{Heliocentric radial velocities and systematic errors for key emission lines at the observed positions as shown in Fig.~\ref{fig9}.}
\begin{tabular}{lccccc}
  \hline
  Obs.     &  \OIII    &    H$\alpha$    &   \NII      &    \SII   & error \\
        &  \tiny{5006.84\AA}  &  \tiny{6562.80\AA}  &  \tiny{6583.41\AA}  &  \tiny{6716.47\AA} &  \\
           &   \tiny{km s$^{-1}$}  &  \tiny{km s$^{-1}$} & \tiny{km s$^{-1}$}  &  \tiny{km s$^{-1}$}   &  \tiny{km s$^{-1}$}   \\
           \hline\hline
  2dF-\tiny{1200R}   & -   & 377 & 391 & 396 & $\pm$\,12\\
  Posn. 1 & 332 & 341 & 376 & 365 & $\pm$\,28 \\
  Posn. 2 & 312 & 390 & 407 & 411 & $\pm$\,28 \\
  \hline
\end{tabular}\label{table3a}
\end{table}


\section{Radio, Infrared and X-ray observations}
\label{section3}

\subsection{Radio observations}

Radio detection has traditionally been the main method for identification and confirmation of SNRs because of their non-thermal spectra, radio shell structures and polarization (e.g., Bozzetto et al. 2014b). Radio observations also alleviate problems associated with optical detection where dust often obscures optical detail. With the development of new instrumentation, the sensitivity of existing radio telescopes is constantly being enhanced, allowing us to improve our knowledge of SNRs within the LMC. For this project we have used a combination of dedicated radio observations complimented by freely available radio survey data.
\begin{figure}
  \includegraphics[width=0.48\textwidth,angle=0]{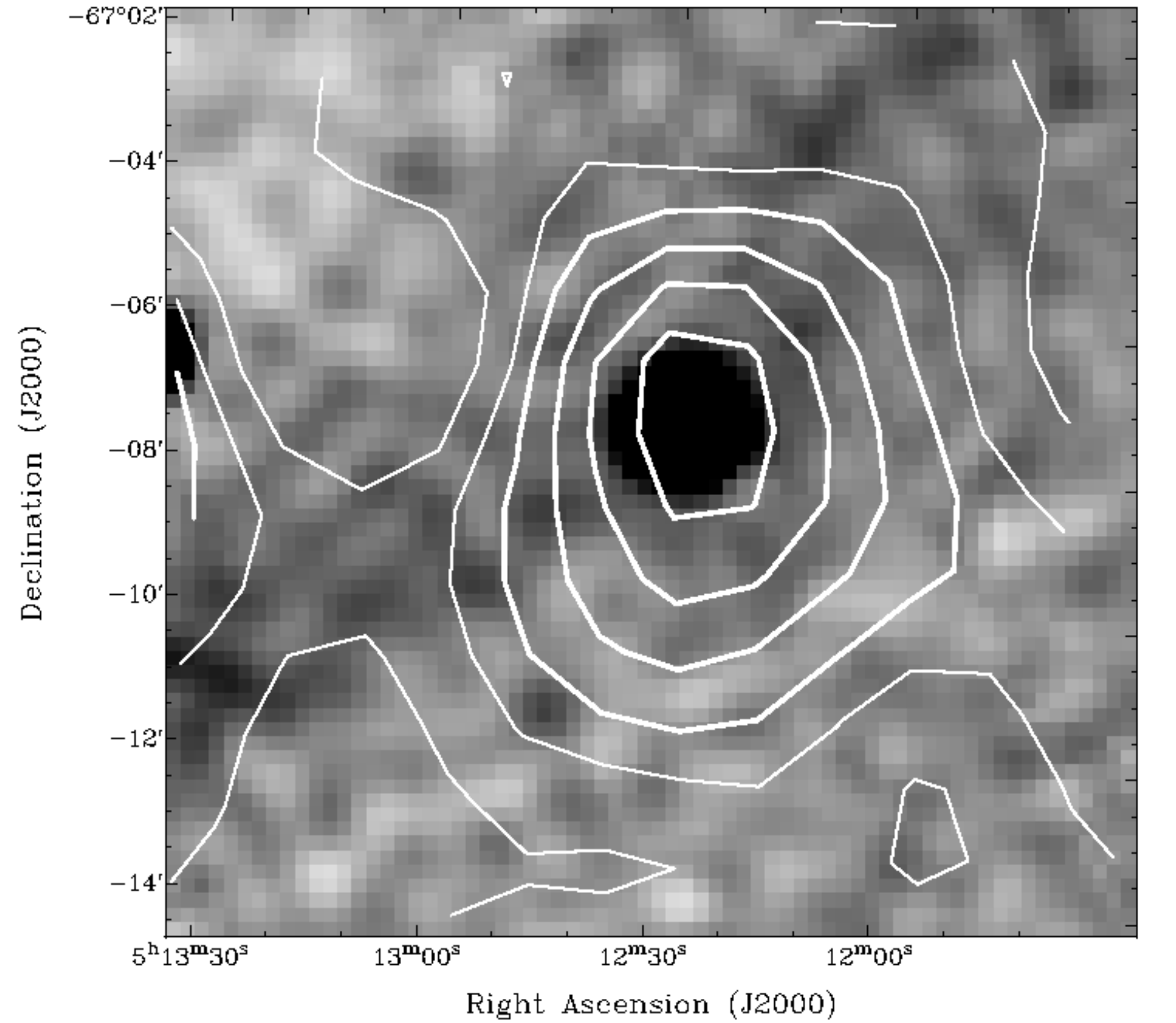}\\
 \caption{Radio contours from PMN at 4850 MHz superimposed on the 843 MHz SUMSS greyscale map of SNR \SNR. Contours are spaced at 0.01\,mJy\,beam$^{-1}$. The flux density varies from --0.66\,mJy to 3.16\,mJy across the area shown.}\label{fig19}
\end{figure}

The radio counterpart to this optically detected SNR was originally found using data from the Parkes-MIT-NRAO (PMN; beam size = 300$^{\prime\prime}$) radio survey at 4\,850\,MHz  (Condon, Griffith, Wright, 1993) and the Sydney University Molonglo Sky Survey (SUMSS; beam size = 45$^{\prime\prime}$) at 843\,MHz (Bock, Large, Sadler, 1999). Although a radio source is detected at the position of \SNR in both of these surveys, it is more extensive in the PMN due to the lower resolution of the survey (see Fig.~\ref{fig19}). The contours therefore do not represent the true size. A detection of \SNR~was made in Filipovi{\'c} et al. (1995; listed as LMC B0512-6710) where a radio continuum analysis of the LMC was performed at various frequencies using the Parkes radio telescope. The SUMMS 843~MHz measurement was omitted from spectral index calculations as it was superseded by the 843~MHz Molonglo Sythesis Telescope (MOST) measurement, due to the higher sensitivity of the MOST image and its greater UV--coverage.

\begin{figure}
  \includegraphics[width=0.48\textwidth,angle=0]{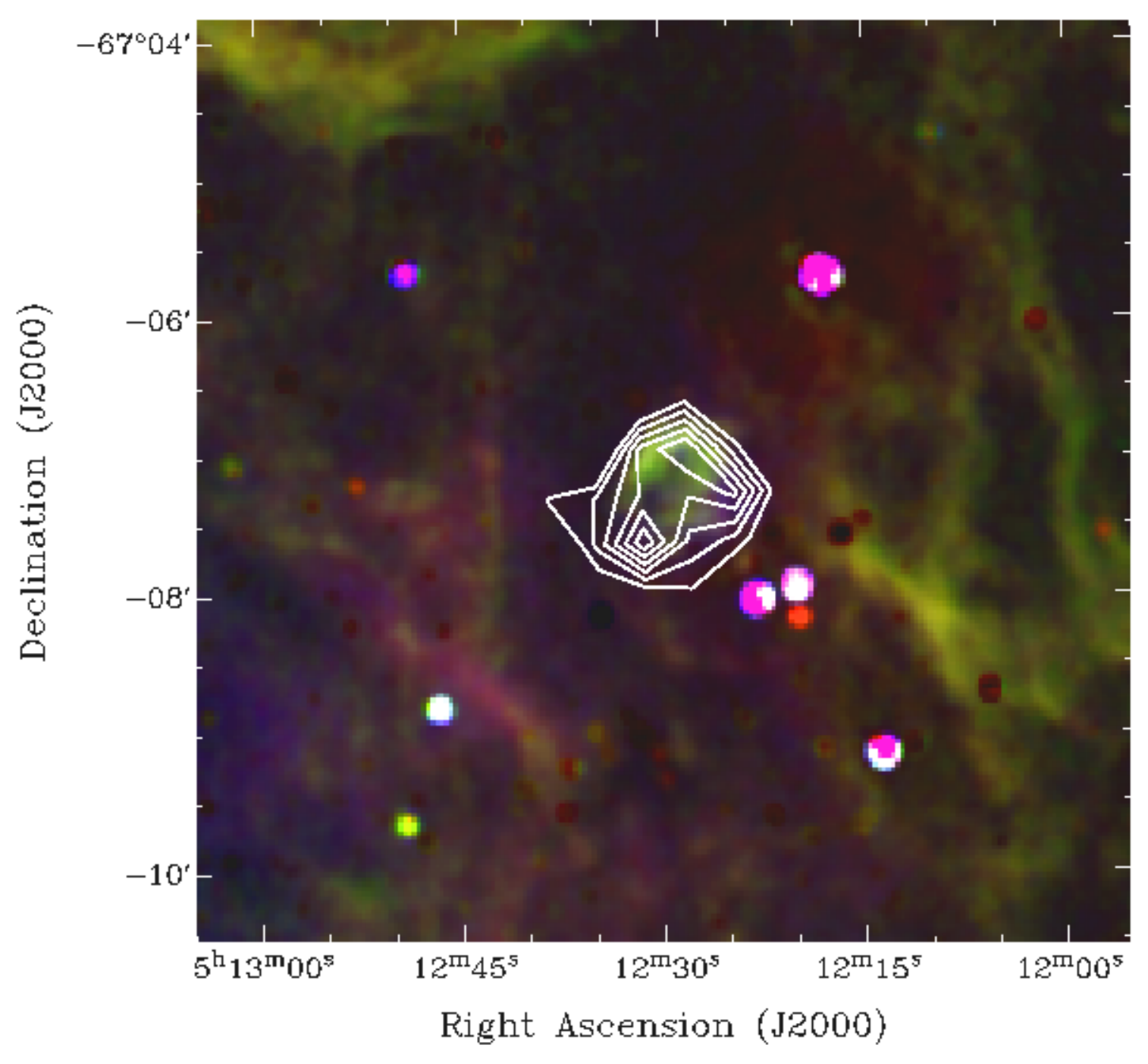}\\
 \caption{\SNR~with contours from the Dickel et al. (2010) 8640\,MHz mosaic image (beam size =\,22\,$\times$\,22\,arcsec). The local rms is $\sigma$\,=\,0.5 mJy\,beam$^{-1}$ and contours are 3, 4, 5, 6, 7 and 8 $\sigma$. The radio contours are superimposed on the continuum-subtracted MCELS map of the immediate area where H$\alpha$ is red, \OIII~is blue and \SII~is green.}\label{fig8a}
\end{figure}

The PMN and SUMSS radio archival data, though helpful in confirmation was of insufficient quality to reveal the structure and so the latest radio-continuum data and our own pointed observations were used to provide details and additional frequencies to demonstrate the non-thermal nature of \SNR. Radio-continuum data used in this project includes a 36~cm (843~MHz) MOST mosaic image (as described in \citealt{1984AuJPh..37..321M}), a 20~cm (1\,377~MHz) mosaic image from \citet{2007MNRAS.382..543H}, and 6~cm (4\,800~MHz) and 3~cm (8\,640~MHz) mosaic images from Dickel et al. (2005, 2010) which are shown as contours against the continuum-subtracted MCELS three-colour map (see section 2) in Fig.~\ref{fig8a}.

\begin{figure}
  \includegraphics[width=0.49\textwidth,angle=0]{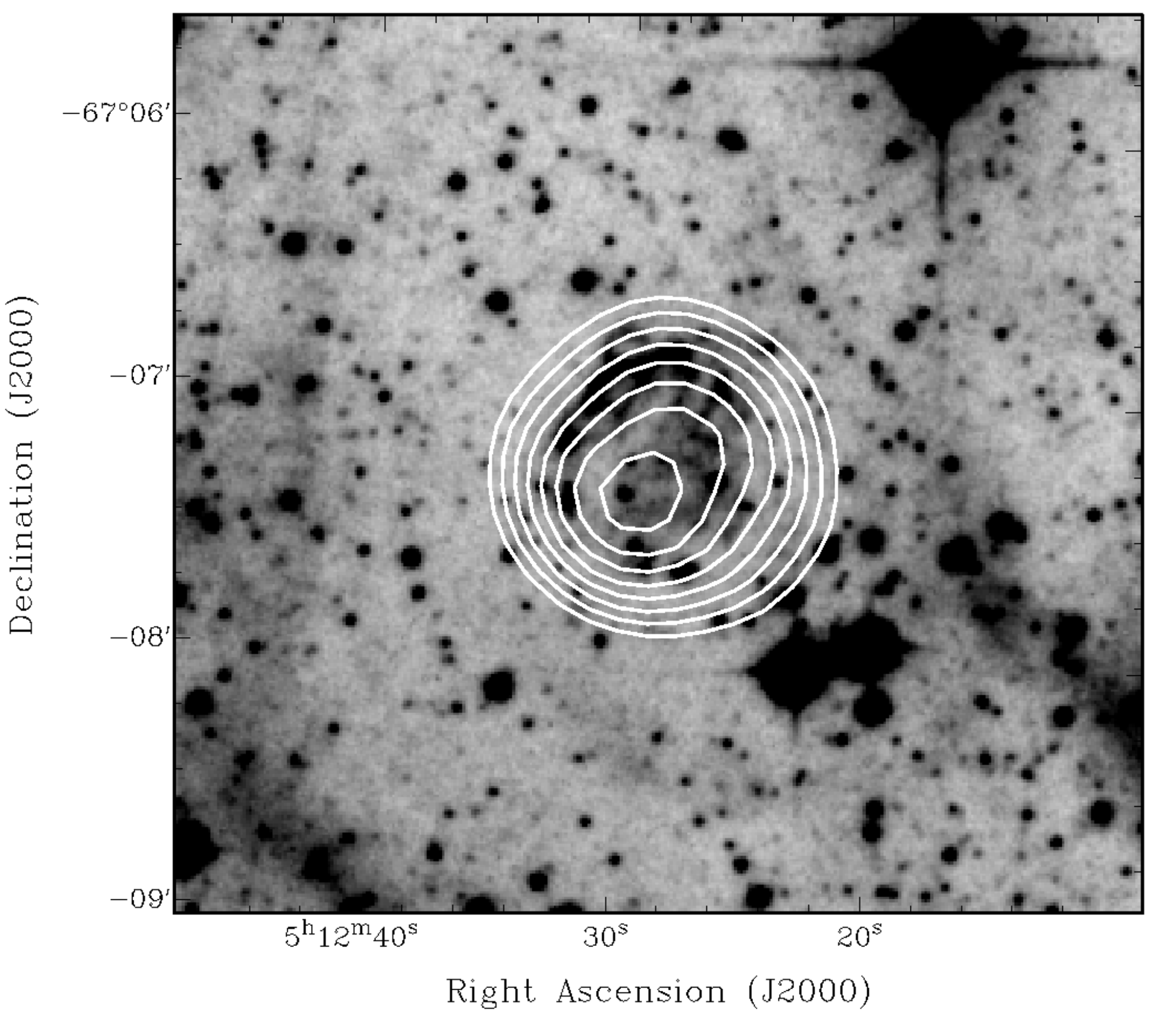}\\
 \caption{Contours using the entire 2 GHz bandwidth centered at 5\,500\,MHz are overlaid on the H$\alpha$ image of \SNR. The contours are spaced at $\sim$\,0.01\,mJy\,beam$^{-1}$ commencing at 0.3\,mJy\,beam$^{-1}$. }\label{fig8b}
\end{figure}

These mosaic data were complemented by pointed observations of \SNR~from our Australian Telescope Compact Array (ATCA) observations taken on 2011 November 15 and 16, using the Compact Array Broadband Backend (CABB) receiver which has a bandwidth of 2~GHz. The compact array configuration EW367 was chosen in an effort to obtain as much emission from the shorter spacings at 6 and 3\,cm ($\nu=5\,500$ and 9\,000\,MHz, respectively) as possible. The resulting 6\,cm contours are overlaid on the MCELS H$\alpha$ map of \SNR~in Fig.~\ref{fig8b}. Total integration time over the two days for this source amounted to $\sim$64\,min. The \textsc{miriad}\footnote{http://www.atnf.csiro.au/computing/software/miriad/}  \citep{1995ASPC...77..433S} and \textsc{karma}\footnote{http://www.atnf.csiro.au/computing/software/karma/} \citep{1995ASPC...77..144G} software packages were used for reduction and analysis. Images were formed using \textsc{miriad} multi-frequency synthesis \citep{1994A&AS..108..585S} and natural weighting. They were deconvolved with primary beam correction applied.

\begin{table*}
\caption{Integrated Flux Densities of \SNR~used in this work.}\label{0512-6707-fluxes-tbl}
\centerline{
\begin{tabular}{cccccl}
\hline
$\nu$	&	$\lambda$	&	rms		&	Flux density		&	Beam Size	&	Image \\
(MHz)	&	(cm)			&	(mJy)	&	(mJy)	&	(arcsec)	&			Ref.		\\
\hline
~\,843		&	36			&	0.8		&	115.5~	&	$46.4\times43.0$		&	MOST		\\
1\,377		&	20			&	1.1		&	86.9		&	$40.0\times40.0$	&	Hughes et al. (2007)			\\
4\,732		&	~6			&	0.2		&	49.6		&	$44.0\times28.4$	&	this work				\\
4\,800		&	~6			&	0.7		&	53.7		&	$35.0\times35.0$	&	Dickel et al. (2010)			\\
5\,244		&	~6			&	0.2		&	47.1		&	$40.1\times25.9$	&	this work			\\
5\,756		&	~6			&	0.2		&	44.3		&	$36.6\times23.7$	&	this work			\\
6\,268		&	~6			&	0.2		&	40.7		&	$33.9\times21.9$	&	this work			\\
8\,640		&	~3			&	0.7		&	29.6		&	$22.0\times22.0$	&	Dickel et al. (2010)			\\
\hline
\end{tabular}}
\end{table*}

\begin{figure}
\centering\includegraphics[angle=-90,scale=0.37]{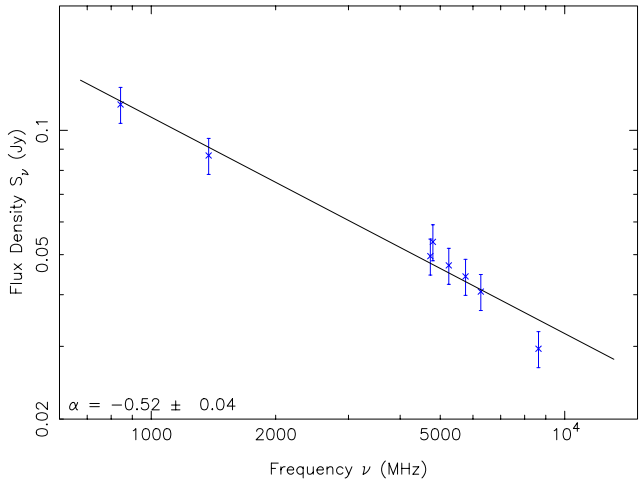}
\caption{Radio-continuum spectral index of \SNR~between 843 and 8\,640\,MHz. The blue markers represent flux density error margins of 10 per cent.\label{rc-spcidx}}
\end{figure}

To estimate the radio spectrum of \SNR, integrated flux density measurements were taken from the four mosaics at frequencies of 843, 1\,377, 4\,800, and 8\,640~MHz, in addition to our observations at 5\,500 and 9\,000~MHz. However, as interferometers such as the ATCA are an arrangement of individual antennae, they lack the short spacings that are responsible for the large-scale emission (e.g., of objects such as SNRs), resulting in a loss in flux density which becomes more pronounced at higher frequencies. For this reason, the 9\,000~MHz integrated flux density measurements were discarded. The 5\,500~MHz observations were split into four 512~MHz bands (4\,732, 5\,244, 5\,756, and 6\,268~MHz), and integrated flux density measurements were taken per band with fitted gaussians using the \textsc{miriad} task \textsc{IMFIT}. The results of these measurements and those taken from the mosaics are listed in Table~\ref{0512-6707-fluxes-tbl} and shown in Fig.~\ref{rc-spcidx}. Errors in these measurements predominately arose from uncertainties in defining the `edge' of the remnant. However, we estimate errors to be less than 10 per cent. The rms noise estimates were taken using the \textsc{miriad} task \textsc{sigest}. The resulting spectral index between 843 and 8\,640~MHz of $\alpha=--0.52\,\pm0.04$ is consistent with the typical SNR spectral index value of $--0.5$.

\subsection{Magnetic field strength}

We use the equipartition formula as given by \citet[and the corresponding `calculator'\footnote{The calculator is available at http://poincare.matf.bg.ac.rs/\~{}arbo/eqp/ }]{2012ApJ...746...79A} to estimate the magnetic field strength of this SNR. However, further to being unable to accurately measure the extent of the remnant, the width of the shell was impossible to measure from a source which is somewhat unresolved in the radio, and consequently, the filling factor was unable to be estimated. Therefore, Table~\ref{tbl:ff} shows the average equipartition field over the whole shell of \SNR~in addition to the estimated minimum energies based on various filling factors ranging from 1 (a completely filled SNR) to 0.25 (a mostly empty shell-like remnant). The range of values for the magnetic field, from $124 - 184~\mu$G, are relatively high for an SNR, and may be indicative of a younger remnant. This can be compared with other LMC SNRs using the same method, e.g., those listed in Table \ref{tbl:equip}. Such values may suggest a remnant of a few thousand years, which is equal to or less than the age estimate of $\sim$2,280 to $\sim$4,690 yrs derived in section~\ref{section2.1}~but still in-keeping with the remnant's relatively small size which ranks it as the 6th smallest currently known in the LMC (Badenes et al. 2010; Bozzetto et al. in prep).

\begin{table}
\caption{Magnetic field strength and minimum energy for varied filling factors \textit{(ff)}.}\label{tbl:ff}
\centerline{
\begin{tabular}{ccc}
\hline
\textit{ff}	&	$B$			&	$E_{min}$				\\
	&	({$\mu$G})	&	(ergs)					\\
\hline
1.00		&	124			&	$6.75\times10^{49}$		\\
0.75		&	134			&	$5.96\times10^{49}$		\\
0.50		&	151			&	$5.01\times10^{49}$		\\
0.25		&	184			&	$3.71\times10^{49}$		\\
\hline
\end{tabular}}
\end{table}

\begin{table}
\caption{Magnetic field strength for a sample of LMC SNRs with their associated ages.}\label{tbl:equip}
\centerline{
\begin{tabular}{crrrl}
\hline
SNR				&	Age~			&	$\Delta$Age	&	$B$			&	Reference\\
				&	(yr)~			&	(yr)			&	({$\mu$G})	&			\\
\hline
0509-6731		&	400*			&	120			&	168			&	\citet{2014b}	\\
0519-6902		&	600*			&	200			&	171			&	\citet{2012a}	\\
0533-7202		&	17\,500~		&	3\,500		&	45			&	\citet{2013}	\\
0508-6902		&	22\,500~		&	2\,500		&	28			&	\citet{2014a}	\\
0529-6653		&	37\,500~		&	12\,500		&	48			&	\citet{2012b}	\\
\hline
\end{tabular}}
* These age estimates are taken from \citet{2005Natur.438.1132R}
\end{table}

\begin{figure}
  \includegraphics[width=0.48\textwidth,angle=0]{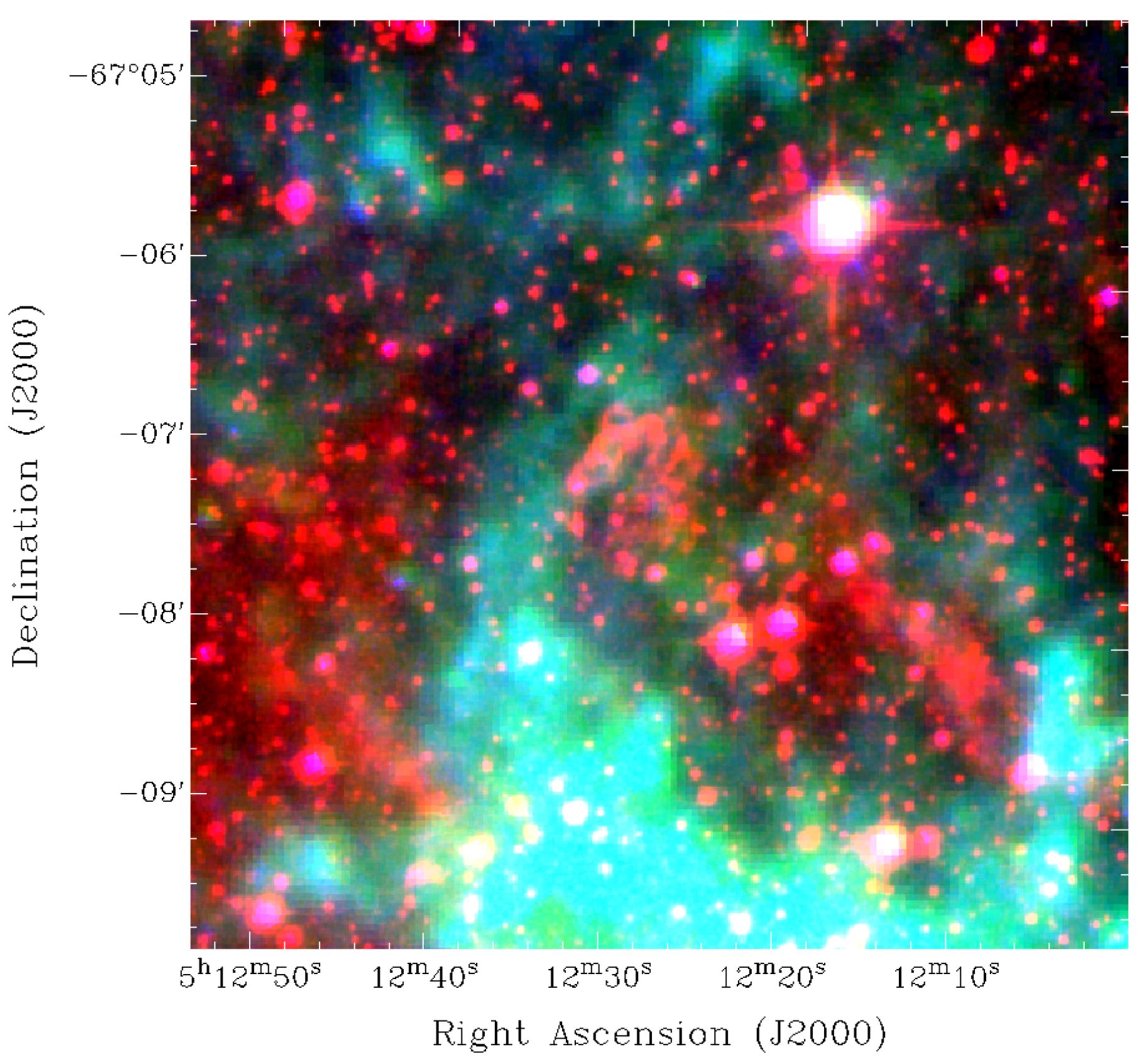}\\
 \caption{The H$\alpha$ map of \SNR~and it's surrounding area (shown in red) is overlaid with the 8$\mu$m SAGE image (blue) and the 24$\mu$m MIPS image (green). }\label{fig23}
\end{figure}

\subsection{Dust}
\label{section3.3}

SNRs heat surrounding dust while mass maps indicate that they also destroy 3.7\,$^{+7.5}_{-2.5}$ M$_{\odot}$ of dust per SNR through the process of sputtering (Laki\'{c}evi\'{c} et al. 2015). Type-II SN are able to produce dust on the order of a solar mass per explosion (Deneault et al. 2003; Nozawa et al. 2003). The average lifetime for interstellar dust under exposed to a SNe are estimated at 2\,$^{+4.0}_{-1.3}$ years (Laki\'{c}evi\'{c} et al. 2015) indicating that the amount of dust that survives may be extremely low. Since SNRs have such a major influence on the local dust environment and effect a galaxy's evolution (Laki\'{c}evi\'{c} et al. 2015) it is worth examining the environment of \SNR~using available MIR imaging. The MIR maps used in this analysis are from ``Surveying the Agents of Galaxy Evolution'' (SAGE) conducted by the Spitzer Space Telescope (Meixner et al. 2006). The 8\,$\mu$m data is from the Infrared Array Camera (IRAC: Fazio et al. 2004) and the 24\,$\mu$m band is from the Multiband Imaging Photometer for Spitzer (MIPS; Rieke et al. 2004). The pixel sizes are 0.6\,$^{\prime\prime}$ for all IRAC wavelengths and 2.49\,$^{\prime\prime}$ for 24\,$\mu$m.

An image from the combined H$\alpha$ (red), 8\,$\mu$m (blue) and 24\,$\mu$m (green) maps (Fig.~\ref{fig23}) shows dense dust emission to the south which is clearly unrelated to the non-thermal SNR. Since the strong dust content at 8\,$\mu$m dominates the other three shorter IRAC wave-bands (see Reid, 2014) they have not been shown in Fig.~\ref{fig23}. The image shows a dust filament that appears to emanate from the strong emission to the south, proceed northwards, along the eastern border of the SNR, and then move across its northern boundary to the west. It therefore appears that the SNR is encased in dust, however, close examination from our line of sight shows a faint, diffuse cloud of 24\,$\mu$m dust across the face of the SNR.

Although the 24\,$\mu$m~emission is spread in patches across the wider area, the contribution from 8\,$\mu$m, shown in blue, is more pronounced in the densest areas of 24\,$\mu$m~dust. The 24\,$\mu$m~dust dominates the region immediately surrounding the SNR, however, patches of 8\,$\mu$m emission are detectable to the east and north. Depending on our viewing angle, it is possible that the expanding SNR has impacted and cleared the 8\,$\mu$m dust towards the SW and along the bright, outer shell at J2000 RA 05$^{h}$ 12$^{m}$ 30$^{s}$ DEC --67$^{\circ}$ 07$^{\prime}$ 55$^{\prime\prime}$ in the North/North-East. A dense region of 8\,$\mu$m dust can be detected south of the southern perimeter of the SNR which is also immediately south of what appears to be the radio center (see Figs.~\ref{fig8a}~\&~\ref{fig8b}).

\subsection{X-ray}

The presence of X-ray emission is typical of most SNRs and is thought to arise from locations where shocks are converting explosive energy into cosmic ray energy. The radiative signatures of GeV and TeV particles indicate radio to X-ray synchrotron emission from relativisitc electrons as well as $\gamma$-ray emission from accelerated electrons and ions (Vink, 2012). The X-ray source corresponding to \SNR~may be found in Haberl \& Pietsch (1999) ROSAT PSPC catalogue of X-ray sources in the LMC region, where it is listed as HP\,483. The central coordinates of the source are J2000 RA 05$^{h}$ 12$^{m}$ 28.0$^{s}$ DEC -67$^{\circ}$ 07\,$^{\prime}$ 27\,$^{\prime}$$^{\prime}$, which is to within $\sim$4 arcsec of the centre of the remnant. The X-ray source extent from the central position is given as 8.5$\pm$0.2 arcsec. The count rate (PSPC 0.1-2.4keV) is given as 3.33$\times$10$^{-3}$\,$\pm$8.7$\times$10$^{-4}$ and the hardness ratios are given as follows: H1 is 1.0\,$\pm$0.38 and H2 is 0.04\,$\pm$0.23. Further X-ray observations will be provided in Kavanagh et al. A\&A, submitted.

\section{Conclusion}
\label{section4}

We have discovered a new SNR in the LMC. This classification is based on combined evidence from optical imaging and spectroscopy, radio detections with a non-thermal spectral index and an X-ray counterpart. Both optical image shown in Fig.~\ref{fig2} clearly depict the shell structure and typical SNR-like morphology. The most distinguishing features are a series of bright loops across the northern boundary of the object. An \SII/H$\alpha$ ratio of 0.93 at one position across the outer shell suggests a typically strong shock. The presence and strength of the \OII\,3727,\,3729 doublet and the \OIII\,4959,\,5007 lines are typical for SNRs. We note that the nitrogen \NII/H$\alpha$ ratios are low compared to those found in the spectra of Galactic SNRs, but this can be attributed to the fact that this element is less abundant in the LMC compared to the Galaxy (see Dopita, 1979). Confusion with Wolf-Rayet ejecta can be ruled out as no WR star was detected in the vicinity and the MCELS survey showed H$\alpha$, not \OIII,~leading the emission. Assuming that the SNR is in the adibatic stage, the Sedov-Taylor method was used to find the dynamical age of the remnant which may be between $\sim$2,000\,yrs and $\sim$5,000\,yrs. The expansion velocity will therefore be between $\sim$\,600\,km s$^{-1}$ and 1400\,km s$^{-1}$.

There is a very strong radio source at the optical center of the SNR. Radio-continuum data used in this project includes 36, 20, 6 and 3\,cm mosaic images. These mosaic data were complemented by pointed observations of \SNR~from our ATCA observations. The results of these measurements and those taken from the mosaics are listed in Table~\ref{0512-6707-fluxes-tbl} and shown in Fig.~\ref{rc-spcidx}. The resulting spectral index between 843 and 8\,640~MHz of $\alpha=-0.52\pm0.04$ is consistent with the typical SNR spectral index value of $-0.5$. Values derived for the magnetic field, from $124 - 184~\mu$G, are relatively high for an SNR, and may suggest an age of only a few thousand years.

This SNR is surrounded by an area filled with dense and often compact dust, much of which may have been expelled outwards or cleared by the expanding shell of the SNR. Any contribution the SNR has made towards the local dust content is impossible to judge due to the large region of dense dust to the immediate south. An X-ray detection at the center of the SNR was made through a previous survey (Haberl \& Pietsch, 1999).

\section*{Acknowledgments}

The authors wish to thank the Australian Astronomical Observatory for observing time on the AAT and the South African Astronomical Observatory for observing time on the 1.9-m telescope. The ATCA is part of the Australia Telescope National Facility which is funded by the Commonwealth of Australia for operation as a National facility managed by the CSIRO. We used the KARMA software package developed by the ATNF. We thank Quentin Parker and Travis Stenborg for observing the SNR during their allocated time on the South African Astronomical Observatory's 1.9-m telescope. WR wishes to thank Macquarie University for a research grant and travel funds.

\appendix






\bsp

\label{lastpage}

\end{document}